\documentclass[12pt,a4paper]{article}
\usepackage[utf8]{inputenc}
\usepackage{amsmath}
\usepackage{amsfonts}
\usepackage{amssymb}
\usepackage{graphicx}

\providecommand{\keywords}[1]{\textbf{Keywords---} #1}

\usepackage[labelformat=simple]{subcaption}

\usepackage{url}

\usepackage{authblk}
\usepackage{color}
\usepackage{epstopdf}
\usepackage{wrapfig}

\usepackage{pgfplots}
\usepackage{tikz}
\usetikzlibrary{shapes}

\usepackage[
square,
sort,
comma,
numbers
]{natbib}

\title{\vspace{-4cm}An efficient, localised approach for the simulation of elastic blood vessels using the lattice Boltzmann method}
\author[1]{McCullough, J.W.S.}
\author[1,2]{Coveney, P.V.}

\affil[1]{Centre for Computational Science, Department of Chemistry, University College London, UK}
\affil[2]{Informatics Institute, University of Amsterdam, Netherlands}

\date{\today}

\begin{document}
\maketitle

\begin{abstract}
Many numerical studies of blood flow impose a rigid wall assumption due to the simplicity of its implementation compared to a full coupling to a solid mechanics model. In this paper, we present a localised method for incorporating the effects of elastic walls into blood flow simulations using the lattice Boltzmann method. We demonstrate that our approach is able to more accurately capture the flow behaviour expected in an elastic walled vessel than a rigid wall model and achieves this without a loss of computational performance. We also demonstrate that our approach can capture trends in wall shear stress distribution captured by fully coupled models in personalised vascular geometries.  \\

\end{abstract}


\section{Introduction}
In many fields of computational biomedicine, in conjunction with the approach of exascale high performance computers, is moving towards the realisation of the virtual human. This concept embodies the creation of a high resolution and highly personalised digital representation of the biophysical processes for a particular individual. Such a model will require scalable computational methods that can study human-scale systems such as a particular organ but can also be efficiently coupled to other such tools build a virtual human model. The development of the virtual human will allow clinicians to determine the optimal course of treatment for a given individual and their personal characteristics.

Numerical simulation of any physical system will require some level of approximation of the fundamental physics involved. A particular examples of this is the assumption of rigid walls in the study of fluid flow through pipe-like structures. Whilst this is valid in mechanical settings for example, organic structures such as blood vessels are known to possess elastic properties. Correct representation of the complex physics of blood flow and vessel behaviour is necessary to develop reliable models for simulating vascular systems. Whether an elastic wall representation is necessary depends on the particular vessels being studied, the scale of the simulation and the flow features of interest. In 1D models, the effect of elastic walls can be straightforwardly implemented through a constitutive relation between pressure and the cross-sectional area of the vessel at solution locations \cite{Sheng1995,Muller2014,Mynard2015b}, the updated area can then inform the flow velocity at those sites. The implementation of elastic walls in a 3D model is significantly more difficult as it typically demands coupling the wall boundaries of the fluid domain to a solid mechanics model for an elastic vessel wall. Depending on the fluid solver used, the changing wall position may demand the fluid domain to be modified in response. Both of these procedures can be complex and computationally costly to conduct in 3D and may be a reason as to why a number of 3D models utilise a rigid wall assumption \cite{Feiger2020,Feiger2021,McCullough2021}. \\

With a view towards the development of the virtual human, we are looking towards conducting full human simulations of arterial and venous vascular geometries using the open-source blood flow simulator HemeLB \cite{HemeLBweb,Mazzeo2008,Bernabeu2012,Bernabeu2014,Nash2014,Groen2018,McCullough2021}. This solver has been specifically optimised to deal with the complex and sparse geometries characteristic of vascular domain. It has demonstrated excellent strong scaling characteristics on such domains on tens and hundreds of thousands of computer cores \cite{Patronis2018,McCullough2021} thanks to the inherent properties of the lattice Boltzmann method (LBM) on which it is based. These properties make HemeLB a good candidate for being able to simulate the blood flow in a virtual human model. \\

In this paper, we introduce a method for representing the effect of elastic walls at the edge of a lattice Boltzmann fluid domain through a modified boundary condition. This approach retains the inherent scalability of the LBM and captures the key features of elastic flow without a loss of performance compared to a rigid wall implementation. The remainder of the paper will discuss the development of our elastic wall model and demonstrate its performance. In Section 2 we will outline the LBM and the boundary conditions relevant to this work. Here we will also compare the results of our model against analytical results and those obtained with a rigid wall assumption. In Section 3 we will use our model to simulate flow in personalised arteries of the left forearm. We discuss findings and provide an outlook for how this work could be incorporated into a full-scale virtual human model in Section 4. \\

\section{Numerical Methods}
In this paper we make use of the LBM to solve 3D flow through a vascular geometry. In this section, we will give a brief outline of this approach followed by a description of our proposed method for replicating the impact of elastic walls. We then verify out implementation through comparison to analytical results that are relevant to blood flow simulation. In particular we will compare against results for Womersley flow in an elastic walled cylinder. This is suitable for comparison in that it resembles both the pulsitility that is characteristic of a heartbeat and explicitly allows for the impact of movable boundaries. \\

\subsection{The lattice Boltzmann method}

Here we will give a brief introduction to the LBM; for a deeper discussion of the technique we refer the reader to the wider literature including textbooks such as \cite{Succi2001,Mohamad2011,Guo2013,Kruger2017,Succi2018}). To describe a flow with the LBM, the domain is partitioned into a Cartesian grid with a constant spacing of $\Delta x$ in all 3D directions. At each nodal location, $\textbf{x}$, a discrete set of values, $f_i(\textbf{x},t)$, is assigned to represent the amount of fluid moving in direction $i$ at time $t$. In this work, we use a D3Q19 model where fluid can stay at the current location or move to one of the 18 neighbours described by the sets: $i = 1-6$ $\left[ \left(\pm 1, 0, 0 \right),\left(0, \pm 1, 0 \right),\left(0, 0, \pm 1 \right) \right)$ and $i = 7-18$ $\left[\left(\pm 1, \pm 1, 0 \right),\left(\pm 1, 0, \pm 1 \right),\left(0, \pm 1, \pm 1 \right) \right]$. The flow described by $f_i(\textbf{x},t)$ evolves over the time step $\Delta t$ with a single relaxation time operator,

\begin{equation}
\label{eq:LBE_SRT}
f_i(\textbf{x}+\textbf{c}_i \Delta t,t + \Delta t) = f_i(\textbf{x},t) -\frac{\Delta t}{\tau}(f_i(\textbf{x},t)-f_i^{eq}(\textbf{x}, t)).
\end{equation}

\noindent Here, $\textbf{c}_i$ indicates the velocity set necessary to move flow to neighbour $i$ in a single time step. $\tau$ relaxes $f_i(\textbf{x},t)$ towards the equilibrium state $f_i^{eq}(\textbf{x}, t)$, a discrete approximation of the Maxwell-Boltzmann distribution. As is demonstrated elsewhere in the literature, a Chapmann-Enskog expansion can be used to demonstrate this framework represents the Navier-Stokes equation for fluid flow in a low Mach number limit. This expansion yields the expansion coefficients, $w_i$, for the equilibrium function, 

\begin{equation}
\label{eq:feq}
f_i^{eq}(\textbf{x}, t) = w_i \rho(\textbf{x},t) \left(1 + \frac{\textbf{c}_i\cdot \textbf{u}}{C_s^2} + \frac{(\textbf{c}_i\cdot \textbf{u})^2}{C_s^4} - \frac{|\textbf{u}|^2 }{C_s^2}\right),
\end{equation}

\noindent For D3Q19 these are 1/3 for $i = 0$ (the source node), 1/18 for $i = 1-6$ and 1/36 for $i = 7-18$. $C_s$ represents the speed of sound of the fluid and evaluates to $\frac{1}{\sqrt{3}}$. Local macroscopic properties of density and momentum can be determined from moments of the $f_i(\textbf{x},t)$ population as,

\begin{equation}
\label{eq:Density}
\rho(\textbf{x},t) = \sum_{i}f_i(\textbf{x}, t),
\end{equation}
\noindent and,
\begin{equation}
\label{eq:Momentum}
\rho(\textbf{x},t)\textbf{u} = \sum_{i}f_i(\textbf{x},t)\textbf{c}_i,
\end{equation}

\noindent respectively. Other relevant physical properties of pressure,
\begin{equation}
\label{eq:Pressure}
p(\textbf{x},t) = C_s^2 \rho(\textbf{x}, t),
\end{equation}
\noindent and viscosity,
\begin{equation}
\label{eq:Viscosity}
\nu = C_s^2 \left(\tau-\frac{1}{2}\right),
\end{equation}
\noindent arise from the Chapmann-Enskog expansion process and associated assumptions.

\subsection{Elastic wall theory}
The analytical work of Womersley \cite{Womersley1955,Womersley1957} represents some of the seminal work in fluid flow through pipes. These focused on idealised results for pulsitile flow through rigid and flexible pipes and have provided a reference case that has been widely used as a verification model for computational fluid mechanics \cite{Nash2014,FigueroaThesis2006,Filonova2020}. For a full derivation we refer to the source works or \cite{FigueroaThesis2006,Filonova2020} but here we highlight equations of particular relevance to the later development of our elastic wall model. \\

These equations describe properties at a given radial, $r$, and axial, $z$, position in time, $t$. The transient pressure field is described by:

\begin{equation}
p(r,z,t) = H e^{i \omega \left(t - \frac{z}{c} \right)} + p_0 + k_s (z-z_0),
\end{equation}
 
\noindent the axial component of velocity is given by the real component of:
\begin{equation}
w(r,z,t) = \frac{k_s}{4 \mu} (r^2 - R^2) + \frac{H}{\rho c} \left[ 1 - M \frac{J_0 \left( \frac{\Lambda r}{R} \right)}{J_0 \left( \Lambda \right)} \right] e^{i \omega \left(t - \frac{z}{c} \right)} ,
\end{equation}

\noindent and the radial component of velocity is given by the real component of:
\begin{equation}
u(r,z,t) = \frac{H i \omega R}{2 \rho c^2} \left[ \frac{r}{R} - M \frac{2 J_1 \left( \frac{\Lambda r}{R} \right)}{ \Lambda J_0 \left( \Lambda \right)} \right] e^{i \omega \left(t - \frac{z}{c} \right)} .
\end{equation}

\noindent In all three expressions, this is a combination of the steady and oscillatory components of flow. In the derivation of these equations, complex variables are used to simplify the expressions associated with the presence of oscillating flow, thus $i=\sqrt{-1}$. The oscillatory component is represented by the term associated with the $e^{i \omega \left(t - \frac{z}{c} \right)}$. $J_n(x)$ represent the $n^{th}$ order Bessel function of the first kind. Within these equations are parameters governed by the vessel and fluid being studied. Here $R$ is the radius of the vessel, $z_0$ is the location of the entry of the cylinder whilst $c$ is the wave speed within the elastic cylinder. $p_0$ represents the static background pressure whilst $k_s$ the static pressure gradient and $H$ represents the amplitude of the oscillatory pressure input. $\omega$ is the frequency of the oscillatory pressure whilst $\Lambda$ is an imaginary version of the Womersley number (= $i^{3/2} \alpha$, where the Womersley number $\alpha = R \sqrt{\frac{\omega \rho}{\mu}}$). For fluid properties, $\rho$ and $\mu$ represent the density and viscosity. $M$ is a elasticity factor derived from vessel and fluid properties. \\

\subsection{Proposed elastic wall boundary condition}
Many existing representations of elastic walls within the LBM require the explicit changing of node types between fluid and solid to represent the change in wall location. Whilst it is possible to take advantage of the inherent locality of LBM, this approach does typically require each lattice site to have a notion of how far away they are from the centre of the vessel. This is easy to achieve in simple representations of blood vessels such as Cartesian aligned cylinders where the coordinates of the node can be used to deduce the local radial position. In patient-specific representations of blood vessels, this can become a much more challenging task as the vessel geometry and orientation can deviate significantly from such simplifying assumptions. Similarly, the geometry and orientation of patient-specific vessels, combined with the often large number of lattice sites representing them, means that pre-computing such radial data is also non-trivial. It is therefore advantageous to have a representation of elastic walls that does not fundamentally rely on knowledge of a site's position within the vessel. \\

From a conceptual point of view, our proposal assumes that the set of LBM fluid nodes represents the minimum fluid volume of the elastic vessel. We then implement a wall boundary condition that provides a non-zero fluid velocity at that location to mimic the effect of the vessel expanding beyond that point. For the generally small changes in vessel diameter \cite{Atkinson2013}, combined with geometrical uncertainty in image-derived vascular simulation models, such an approach provides a useful compromise between simple rigid wall modelling and the complex and computationally expensive coupling to a solid mechanics model to capture qualitative effects of elastic walled vessels. \\

In our implementation, we build upon the Guo, Zheng and Shi \cite{Guo2002a} (GZS) wall boundary condition. This method was selected due to its basis as an extrapolation condition - a similar concept of what we are trying to achieve. The GZS scheme was proposed as a method for representing curved boundaries. The non-equilibrium component of the distribution at the wall node is constructed from the neighbouring fluid node whilst the equilibrium component is constructed based on the desired location and characteristics of the curved boundary. For cases where the curved boundary is close to the wall (where $\Delta$, being the fraction of the unit cell the boundary is from the fluid node, is $\geq 0.75$), the wall node velocity is proposed by GZS to be $U_w = (U_{boundary} + (\Delta - 1) U_f)/\Delta$. This is used to construct $f_i^{eq}(\textbf{x}_w, t)$ whilst $f_i^{neq}(\textbf{x}_w, t)$ is taken to be the same as $f_i^{neq}(\textbf{x}_f, t)$ in the post-collision construction of $f_i(\textbf{x}_w, t) = f_i^{eq}(\textbf{x}_w, t) + (1 - \tau^{-1}) f_i^{neq}(\textbf{x}_w, t)$.\\

\begin{figure}[!ht]
   \centering
   \includegraphics[width=0.98\textwidth]{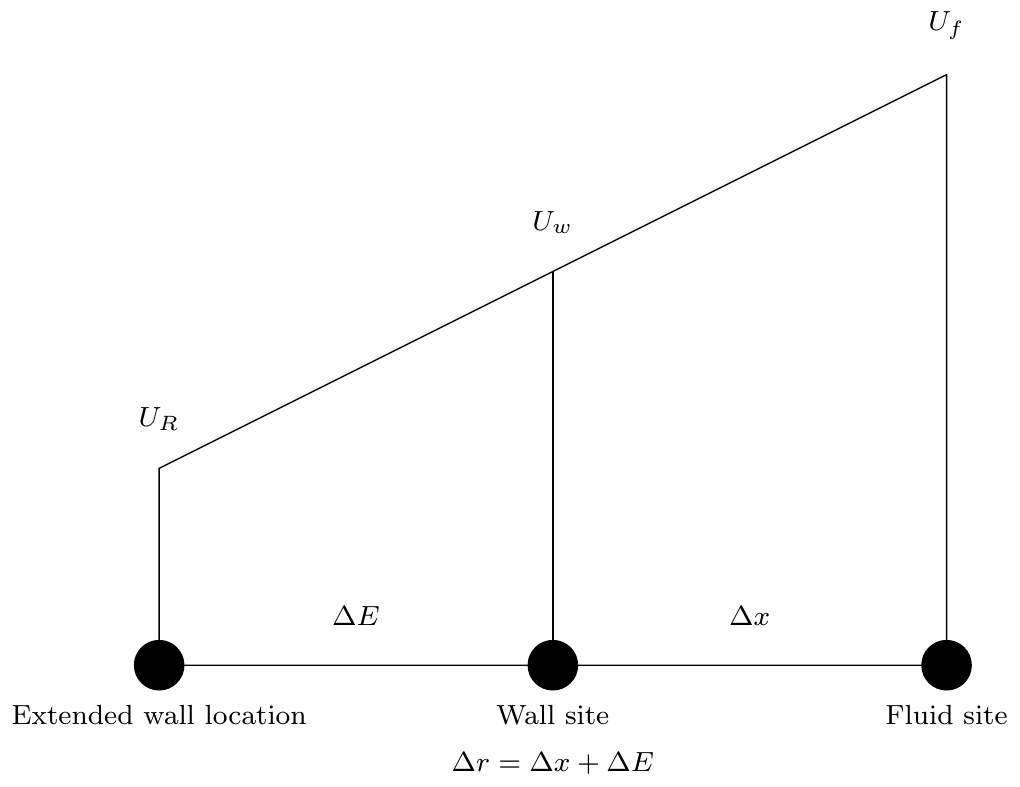}
   \caption{Schematic layout of fluid node, wall node and hypothetical wall location for the proposed boundary scheme.}
   \label{VelScheme}
\end{figure}

In our boundary condition implementation, we consider the case of $\Delta = 1$ to apply a non-zero velocity at the wall node that is approximated to replicate the effect of an elastic boundary stretching beyond this point. The conceptual layout of this is given in Figure \ref{VelScheme}. 

If it is assumed that there is a linear change between $U_R$ and $U_f$, then $U_w = U_R + \frac{\Delta r - \Delta x}{\Delta r}( U_f - U_R)$. The $\Delta r$ term can be computed using common relations between pressure and the stiffness of the elastic wall. This expression for $U_w$ can then be fully derived in terms of the known value $U_f$ but taking into account analytic expressions elastic vessels. Here we particularly look for the ratio between the edge of the elastic walled pipe and the location of the fluid node $\Delta E$ inside of it. For a ratio of $F = \frac{U_R}{U_f}$, the value of $U_w$ can be solely computed from $U_f$ as $U_w = \frac{F + \Delta r - \Delta x}{\Delta r}U_f$. For this computation we will only consider the oscillating part of the axial component of flow. At the extended wall location, the axial velocity is:

\begin{equation}
w(R,z,t) = \frac{H}{\rho c} \left[ 1 - M \right] e^{i \omega \left(t - \frac{z}{c} \right)} ,
\end{equation}

\noindent whilst at the location of the fluid node it is:

\begin{equation}
w(R - \Delta r,z,t) = \frac{H}{\rho c} \left[ 1 - M \frac{J_0 \left( \frac{\Lambda (R- \Delta r)}{R} \right)}{J_0 \left( \Lambda \right)} \right] e^{i \omega \left(t - \frac{z}{c} \right)}.
\end{equation}

\noindent Combining these two expressions we get the following for the ratio $F$:

\begin{equation}
F = \frac{ 1 - M }{ 1 - M \frac{J_0 \left( \frac{\Lambda (R-\Delta r)}{R} \right)}{J_0 \left( \Lambda \right)} } = \frac{ 1 - M }{ 1 - M \frac{J_0 \left( \Lambda (1 - \frac{d\Delta r}{R} ) \right)}{J_0 \left( \Lambda \right)} }.
\end{equation}

The value of F is dependent on the Womersley number ($\alpha$) of the local flow through $\Lambda$ and the extension of the flow $\Delta r/R$. $\alpha$ varies widely throughout the human vascular system from $O(10^{-3})$ in the capillaries to $O(10)$ in the aorta. In our simulation efforts, the resolution of available human-scale domains means that we typically consider relatively large vessels where $\alpha>1$. Equally, within blood vessels, the amount of flow induced radial dilation is typically a relatively small value and often less than 10\% of the radius \cite{Atkinson2013}. Whilst $M$ also varies based on vessel radius and Womersley number, its variation is much less than that of $\Lambda$ and $\Delta E$. \\

\begin{figure}[!ht]
   \centering
   \includegraphics[width=0.98\textwidth]{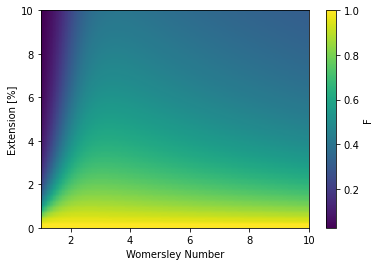}
   \caption{Distribution of $F$ for the range of physiologically relevant values of $\Delta E$ and Womersley number in vessels of radius 1mm.}
   \label{FHeatMap}
\end{figure}

Based on these variations of parameters, we can generate a map of values of $F$ for human vessels to apply for a given simulation. In cases where the variation in vessel characteristics is relatively small, this can help to narrow the selected value of F from this map. We precompute $F$ for our boundary condition as the local Womersley number may not be known for a given boundary location within a large vascular tree. In Figure \ref{FHeatMap} we illustrate the distribution of values of $F$ for vessels of radius 1mm. \\

\subsection{Model verification}

In \cite{FigueroaThesis2006,Filonova2020}, the problem of flow through the carotid artery is examined as a test case for an elastic wall model. Here we demonstrate how our model is able to capture the essence of elastic wall flow whilst not losing computational performance compared to a rigid wall model. Here we verify the performance of our model through the comparison to the analytical solutions for Womersley flow in an elastic vessel with a particular focus on the oscillatory component of flow. We consider a cylinder of $R$ = 3~mm and total length of 4~cm. The vessel walls are set at a thickness of $h$ = 0.1$R$ and have material properties of: Young's modulus $Y$ = 20~kPa, Poisson ratio $\sigma$ = 0.5 and density of 1000 kg/m$^3$. We assume that the fluid has a density of 1000~kg/m$^3$ and viscosity of 0.004~Pa.s. The applied oscillating pressure gradient has a period of $\frac{\pi}{2}$ seconds. The flow corresponds to a Womersley number of 3.0. We link the expansion of the vessel to the pressure via $\Delta E = \frac{(1-\sigma^2)R^2}{Y h}(p - P_0)$, where $P_0$ is the pressure at which $\Delta E = 0$. We used three different levels of resolution to study this case - $R = 50 \Delta x$, $100 \Delta x$ and $200 \Delta x$ as well as two levels of applied pressure gradient amplitude - $k_p = -50 Pa/m$ and $-150 Pa/m$. We compare the numerical results to those obtained from the analytical solutions. In particular we will examine the axial velocity obtained along the centreline of the cylindrical test domain and across the radius at a plane in the centre of the domain. As the radius of the cylinder is $3mm$, and the extension was observed to be relatively small, we will choose a value for $F$ of 0.85. All simulations were conducted on the SuperMUC-NG supercomputer situated at the Leibniz Supercomputing Centre, Germany. We provide details on the computational configurations used for our simulations in the Appendix. \\

\begin{figure}[!ht]
 \begin{subfigure}{0.49\textwidth}
   \centering
   \includegraphics[width=0.98\textwidth]{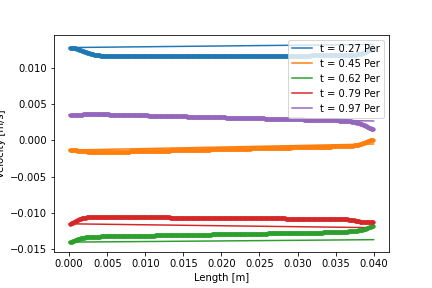} 
  \caption{$R = 50\Delta x$ velocity}
  \label{R50k150VelocityL}
 \end{subfigure}
  \begin{subfigure}{0.49\textwidth}
   \centering
   \includegraphics[width=0.98\textwidth]{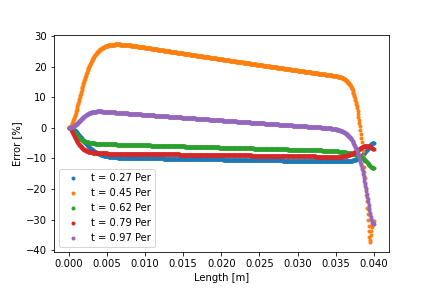} 
  \caption{$R = 50\Delta x$ error}
  \label{R50k50ErrorL}
 \end{subfigure}
 \begin{subfigure}{0.49\textwidth}
   \centering
   \includegraphics[width=0.98\textwidth]{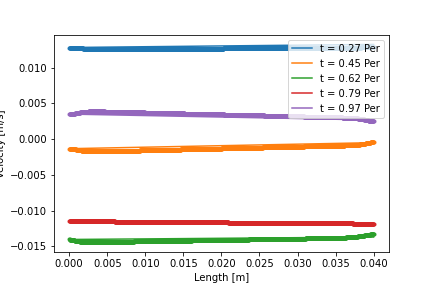} 
  \caption{$R = 100\Delta x$ velocity}
  \label{R100k50VelocityL}
 \end{subfigure}
  \begin{subfigure}{0.49\textwidth}
   \centering
   \includegraphics[width=0.98\textwidth]{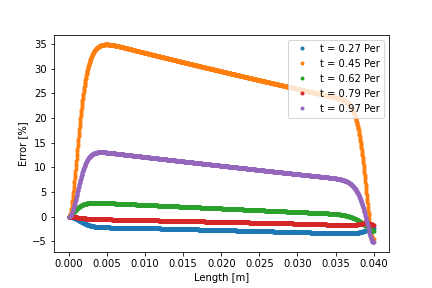} 
  \caption{$R = 100\Delta x$ error}
  \label{R100k50ErrorL}
 \end{subfigure}
  \begin{subfigure}{0.49\textwidth}
   \centering
   \includegraphics[width=0.98\textwidth]{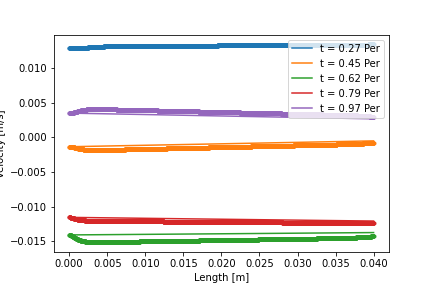} 
  \caption{$R = 200\Delta x$ velocity}
  \label{R200k50VelocityL}
 \end{subfigure}
  \begin{subfigure}{0.49\textwidth}
   \centering
   \includegraphics[width=0.98\textwidth]{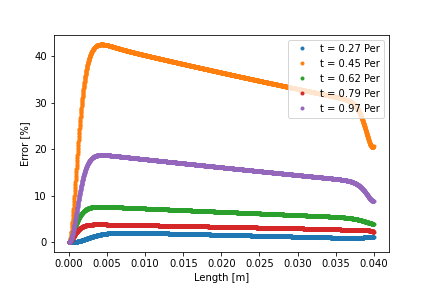} 
  \caption{$R = 200\Delta x$ error}
  \label{R200k50ErrorL}
 \end{subfigure}
  \caption{Velocity profiles and relative error along the central axis of the test cylinder comparing the current elastic wall (dots) to the elastic wall analytical solution (solid lines).}
  \label{ElasticLength_k50}
\end{figure}

In Figures \ref{ElasticPlane_k50} and \ref{ElasticPlane_k150} we compare the central plane velocity profiles and observed relative error for the two pressure gradient cases at each of the examined geometric resolutions. In our results, we allowed the simulation to overcome initialisation effects and then compared the calculated profiles at 5 stages within an oscillation period. Generally speaking, our model is able to resolve the expected analytic results with less than 10\% error, with the greatest error being observed at the time steps with the lowest flow velocity magnitudes in the central plane where relative errors can be magnified. It can also be noted that there is varying error behaviour as the resolution of the cylinder is increased, this may be related to the choice of $F$ providing a better approximation of the elastic wall flows in some circumstances. When simulations were stable, very similar error trends were observed at higher values of $k_p$ that we tested but do not report here. In Figure \ref{ElasticLength_k50}, we demonstrate similar trends for the axial velocity recorded along the central axis of the cylinder for the case of $k_p = -50 Pa/m$.\\

\begin{figure}[!ht]
 \begin{subfigure}{0.49\textwidth}
   \centering
   \includegraphics[width=0.98\textwidth]{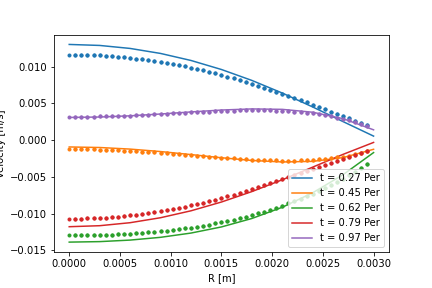} 
  \caption{$R = 50\Delta x$ velocity}
  \label{R50k50Velocity}
 \end{subfigure}
  \begin{subfigure}{0.49\textwidth}
   \centering
   \includegraphics[width=0.98\textwidth]{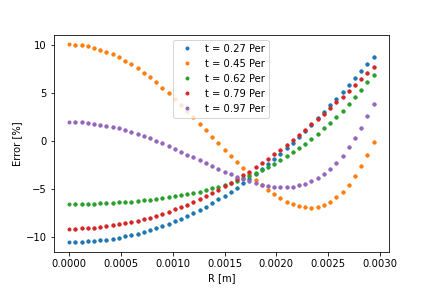} 
  \caption{$R = 50\Delta x$ error}
  \label{R50k50Error}
 \end{subfigure}
 \begin{subfigure}{0.49\textwidth}
   \centering
   \includegraphics[width=0.98\textwidth]{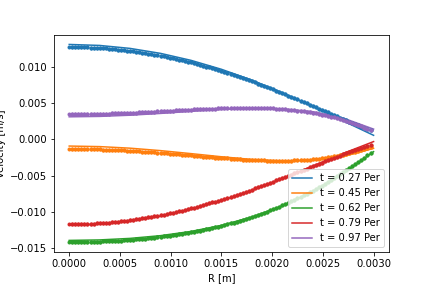} 
  \caption{$R = 100\Delta x$ velocity}
  \label{R100k50Velocity}
 \end{subfigure}
  \begin{subfigure}{0.49\textwidth}
   \centering
   \includegraphics[width=0.98\textwidth]{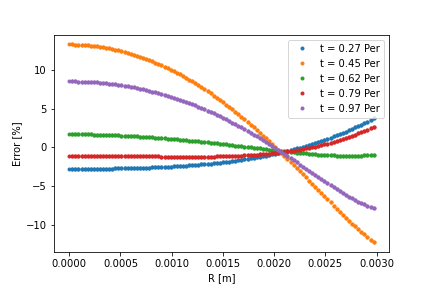} 
  \caption{$R = 100\Delta x$ error}
  \label{R100k50Error}
 \end{subfigure}
  \begin{subfigure}{0.49\textwidth}
   \centering
   \includegraphics[width=0.98\textwidth]{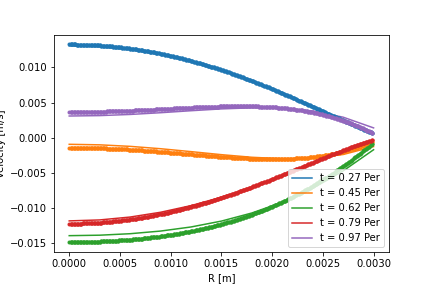} 
  \caption{$R = 200\Delta x$ velocity}
  \label{R200k50Velocity}
 \end{subfigure}
  \begin{subfigure}{0.49\textwidth}
   \centering
   \includegraphics[width=0.98\textwidth]{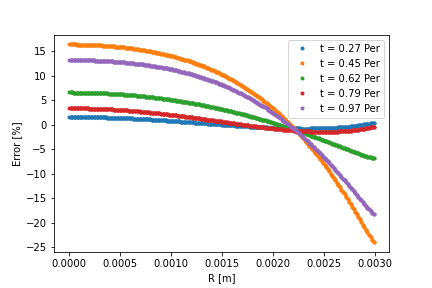} 
  \caption{$R = 200\Delta x$ error}
  \label{R200k50Error}
 \end{subfigure}
  \caption{Velocity profiles and relative error at the centre plane of the test cylinder comparing the current elastic wall model (dots) to the elastic wall analytical equations (solid lines) - $k_p = -50 Pa/m$.}
  \label{ElasticPlane_k50}
\end{figure}

\begin{figure}[!ht]
 \begin{subfigure}{0.49\textwidth}
   \centering
   \includegraphics[width=0.98\textwidth]{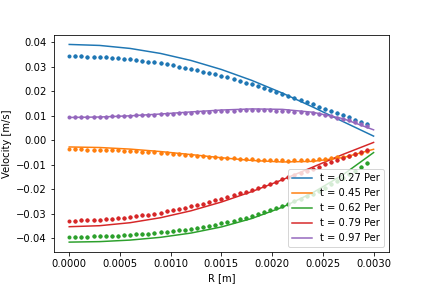} 
  \caption{$R = 50\Delta x$ velocity}
  \label{R50k150Velocity}
 \end{subfigure}
  \begin{subfigure}{0.49\textwidth}
   \centering
   \includegraphics[width=0.98\textwidth]{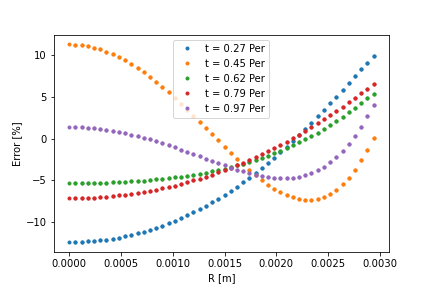} 
  \caption{$R = 50\Delta x$ error}
  \label{R50k150Error}
 \end{subfigure}
 \begin{subfigure}{0.49\textwidth}
   \centering
   \includegraphics[width=0.98\textwidth]{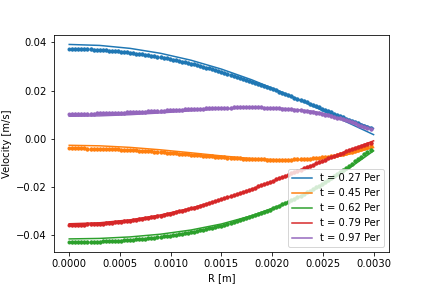} 
  \caption{$R = 100\Delta x$ velocity}
  \label{R100k150Velocity}
 \end{subfigure}
  \begin{subfigure}{0.49\textwidth}
   \centering
   \includegraphics[width=0.98\textwidth]{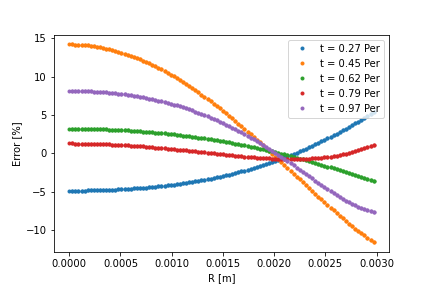} 
  \caption{$R = 100\Delta x$ error}
  \label{R100k150Error}
 \end{subfigure}
  \begin{subfigure}{0.49\textwidth}
   \centering
   \includegraphics[width=0.98\textwidth]{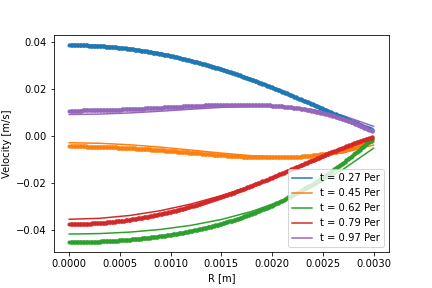} 
  \caption{$R = 200\Delta x$ velocity}
  \label{R200k150Velocity}
 \end{subfigure}
  \begin{subfigure}{0.49\textwidth}
   \centering
   \includegraphics[width=0.98\textwidth]{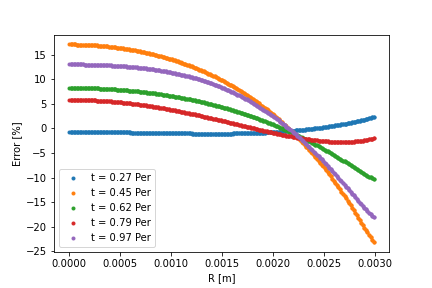} 
  \caption{$R = 200\Delta x$ error}
  \label{R200k150Error}
 \end{subfigure}
  \caption{Velocity profiles and relative error at the centre plane of the test cylinder comparing the current elastic wall model (dots) to the elastic wall analytical solution (solid lines) - $k_p = -150 Pa/m$.}
  \label{ElasticPlane_k150}
\end{figure}

\begin{figure}[!ht]
 \begin{subfigure}{0.49\textwidth}
   \centering
   \includegraphics[width=0.98\textwidth]{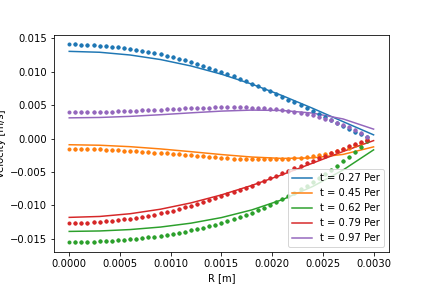} 
  \caption{$R = 50\Delta x$ velocity}
  \label{R50NoEVelocity}
 \end{subfigure}
  \begin{subfigure}{0.49\textwidth}
   \centering
   \includegraphics[width=0.98\textwidth]{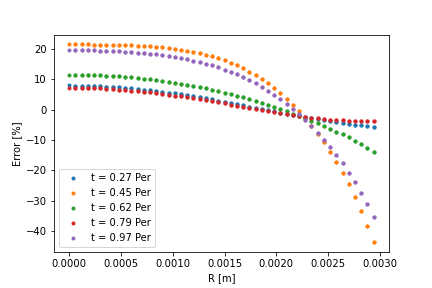} 
  \caption{$R = 50\Delta x$ error}
  \label{R50NoEError}
 \end{subfigure}
 \begin{subfigure}{0.49\textwidth}
   \centering
   \includegraphics[width=0.98\textwidth]{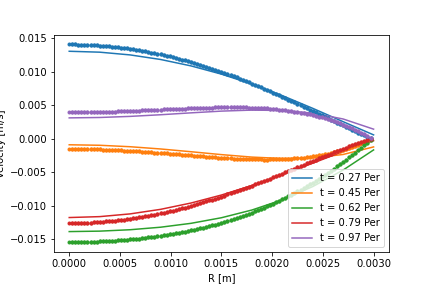} 
  \caption{$R = 100\Delta x$ velocity}
  \label{R100NoEVelocity}
 \end{subfigure}
  \begin{subfigure}{0.49\textwidth}
   \centering
   \includegraphics[width=0.98\textwidth]{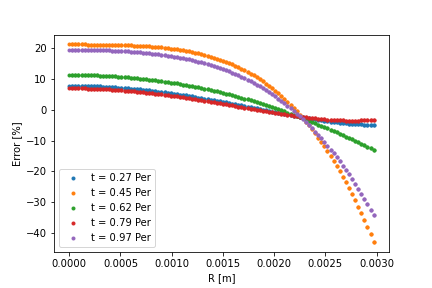} 
  \caption{$R = 100\Delta x$ error}
  \label{R100NoEError}
 \end{subfigure}
  \begin{subfigure}{0.49\textwidth}
   \centering
   \includegraphics[width=0.98\textwidth]{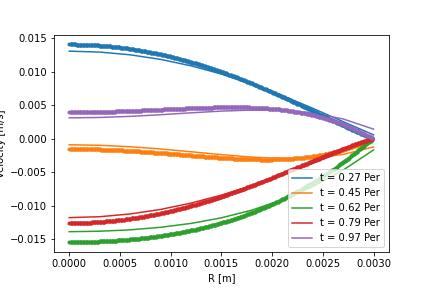} 
  \caption{$R = 200\Delta x$ velocity}
  \label{R200NoEVelocity}
 \end{subfigure}
  \begin{subfigure}{0.49\textwidth}
   \centering
   \includegraphics[width=0.98\textwidth]{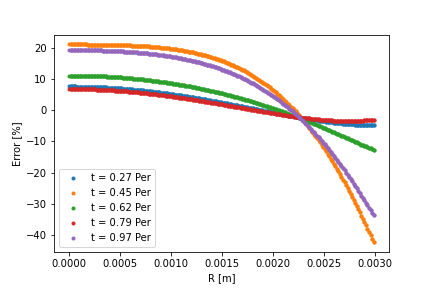} 
  \caption{$R = 200\Delta x$ error}
  \label{R200NoEError}
 \end{subfigure}
  \caption{Velocity profiles and relative error at the centre plane of the test cylinder comparing a rigid wall assumption (dots) to the elastic wall analytical solution (solid lines). Note the significantly larger error at the walls ($R=0.003m$)}
  \label{RigidPlane}
\end{figure}

\begin{figure}[!ht]
 \begin{subfigure}{0.49\textwidth}
   \centering
   \includegraphics[width=0.98\textwidth]{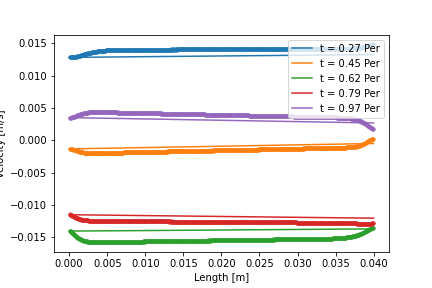} 
  \caption{$R = 50\Delta x$ velocity}
  \label{R50NoEVelocityL}
 \end{subfigure}
  \begin{subfigure}{0.49\textwidth}
   \centering
   \includegraphics[width=0.98\textwidth]{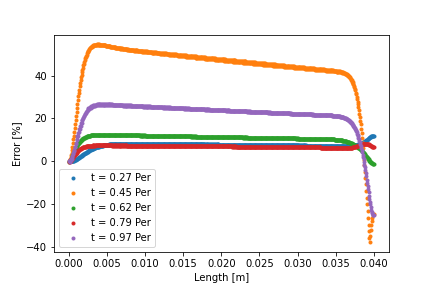} 
  \caption{$R = 50\Delta x$ error}
  \label{R50NoEErrorL}
 \end{subfigure}
 \begin{subfigure}{0.49\textwidth}
   \centering
   \includegraphics[width=0.98\textwidth]{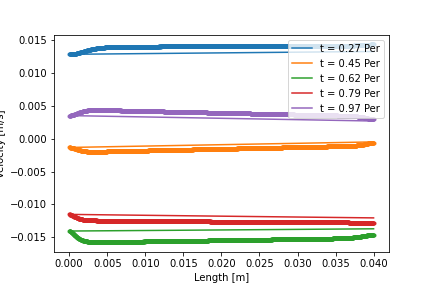} 
  \caption{$R = 100\Delta x$ velocity}
  \label{R100NoEVelocityL}
 \end{subfigure}
  \begin{subfigure}{0.49\textwidth}
   \centering
   \includegraphics[width=0.98\textwidth]{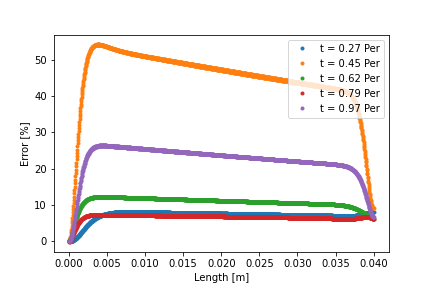} 
  \caption{$R = 100\Delta x$ error}
  \label{R100NoEErrorL}
 \end{subfigure}
  \begin{subfigure}{0.49\textwidth}
   \centering
   \includegraphics[width=0.98\textwidth]{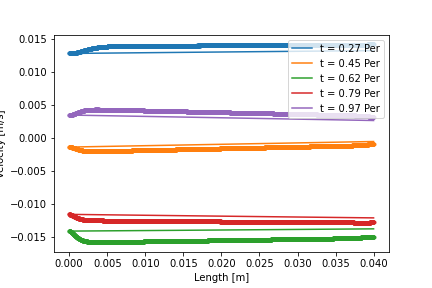} 
  \caption{$R = 200\Delta x$ velocity}
  \label{R200NoEVelocityL}
 \end{subfigure}
  \begin{subfigure}{0.49\textwidth}
   \centering
   \includegraphics[width=0.98\textwidth]{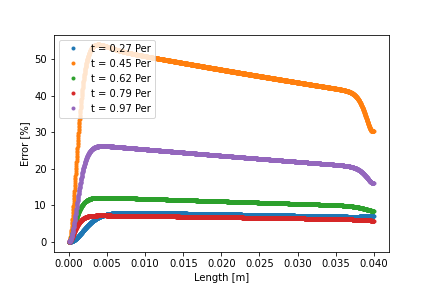} 
  \caption{$R = 200\Delta x$ error}
  \label{R200NoEErrorL}
 \end{subfigure}
  \caption{Velocity profiles and relative error along the central axis of the test cylinder comparing the rigid wall model (dots) to the elastic wall analytical solution (solid lines).}
  \label{RigidLength}
\end{figure}

For the case of $k_p=-50Pa/m$ we have also generated results when a rigid wall assumption is in place. Here we use the well-known LBM bounceback condition to represent the solid walls. Velocity flow profiles at the central plane are presented in Figure \ref{RigidPlane} whilst comparison to the central axis velocity is provided in Figure \ref{RigidLength}. In both of these cases, the error observed when rigid walls are enforced is notably greater than that seen with the our proposed elastic wall condition. These collective results indicate that we are able to capture the key flow results associated with an elastic wall better using our model than can be achieved with a rigid wall implementation. \\

A rigorous implementation of an elastic wall model would typically require the coupling of a solid mechanics solver that moves the wall in response to pressure from the fluid and in return modifies the fluid flow domain. Implementation of this can be time consuming in terms of both computational development and implementation as well as execution time. For this reason, a number of studies will fall back on a rigid wall assumption due to its simplicity  \cite{Feiger2020,Feiger2021,McCullough2021}. In Table \ref{tab:WallTimes} we compare the average time taken to complete 1000 iterations for our elastic wall implementation and the same domain and core configuration using the rigid wall assumption. As can be seen, the elastic wall performance is actually slightly superior to the rigid wall model for the implementations present within our code. Combined with the comparison to the numerical results for our model and the rigid wall implementation above, our model is better able to capture flow characteristics of elastic vessels better than when a rigid wall model is used with no loss of computational performance. This demonstrates that our model can overcome the arguments of the using a rigid wall model due to its simplicity despite the loss of physical accuracy of elastic walls. \\

\begin{table}
\caption{Performance of elastic and rigid wall models - walltime for 1000 iterations}
\begin{center}
\begin{tabular}{|c|c|c|}
\hline 
\textbf{Radius} & \textbf{Elastic [s]} & \textbf{Rigid [s]} \\  
\hline 
$ 50\Delta x$ & 0.9770 & 0.9781 \\ 
\hline 
$100\Delta x$ & 3.159 & 3.385 \\ 
\hline
$200\Delta x$ & 10.656 & 11.446 \\ 
\hline
\end{tabular} 
\end{center}
\label{tab:WallTimes}
\end{table}

\section{Model Application}
\begin{figure}[!ht]
 \begin{subfigure}{0.49\textwidth}
   \centering
   \includegraphics[width=0.98\textwidth]{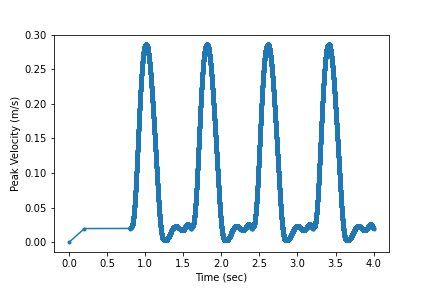} 
  \caption{Arterial inlet profile}
  \label{InletVelocity}
 \end{subfigure}
  \begin{subfigure}{0.49\textwidth}
   \centering
   \includegraphics[width=0.98\textwidth]{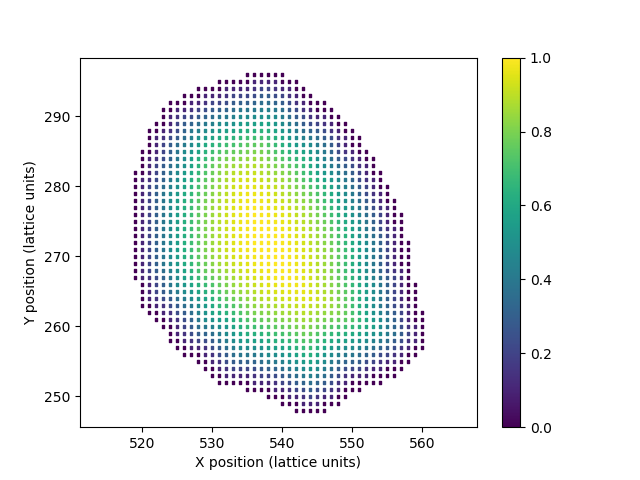} 
  \caption{Arterial inlet distribution}
  \label{InletDistribution}
 \end{subfigure}
 \begin{subfigure}{0.98\textwidth}
   \centering
   \includegraphics[width=0.98\textwidth]{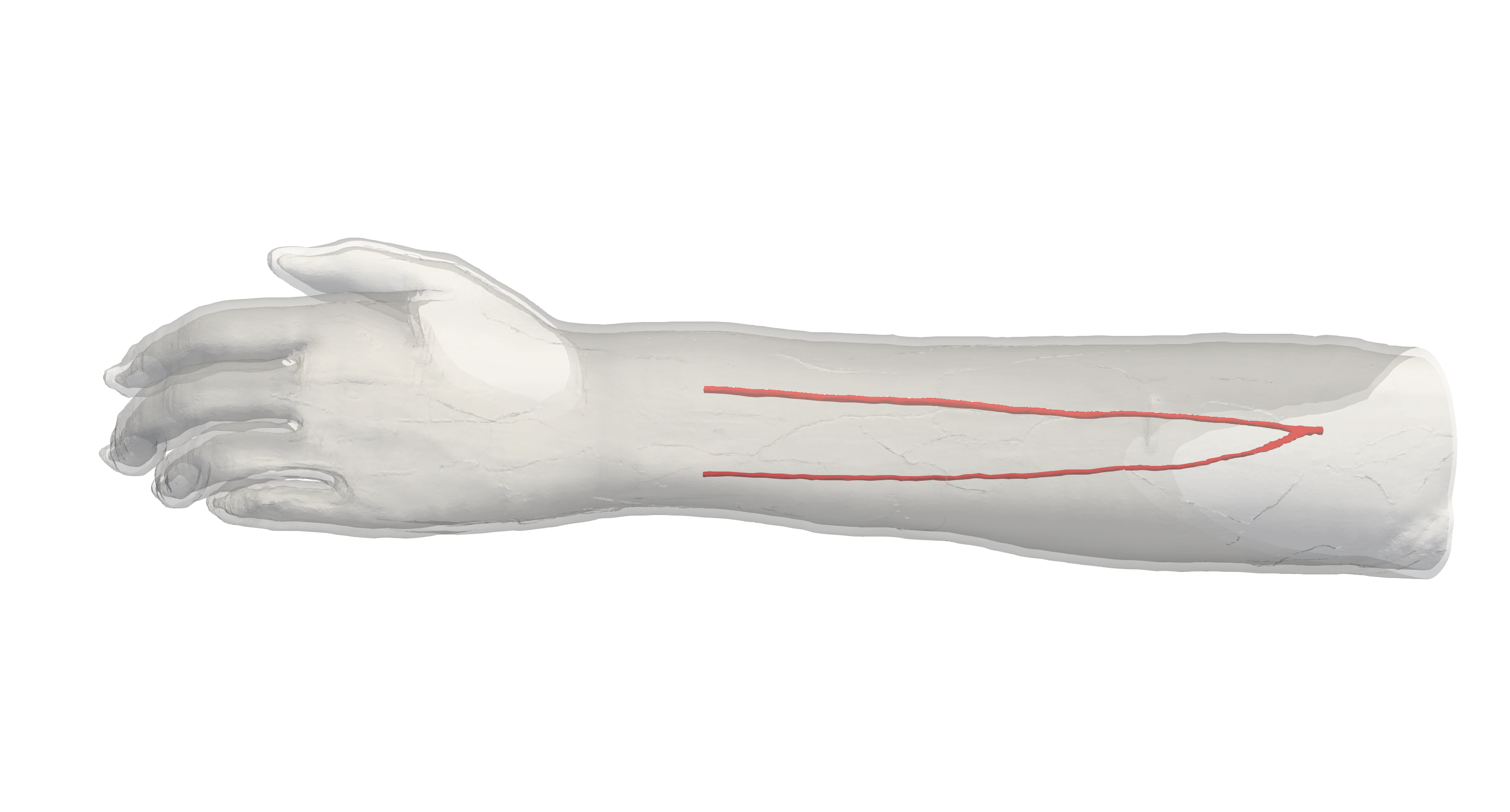}
   \caption{Test vessels}
   \label{Layout}
 \end{subfigure}
  \caption{Simulation inlet conditions used in for the arterial studies. \subref{InletVelocity} details the flow velocity provided to the inlet of the arterial geometry for the forearm flow cases. The first 0.8s of the flow represents an initial warm-up period of flow within the system.\subref{InletDistribution} indicates the distribution of scaling weights applied to the flow velocity at the inlet of the arterial geometry. \subref{Layout} illustrates the flow domain itself within the left forearm.}
  \label{ArteriesLayout}
\end{figure}

\begin{figure}[!ht]
 \begin{subfigure}{0.49\textwidth}
   \centering
   \includegraphics[width=0.98\textwidth]{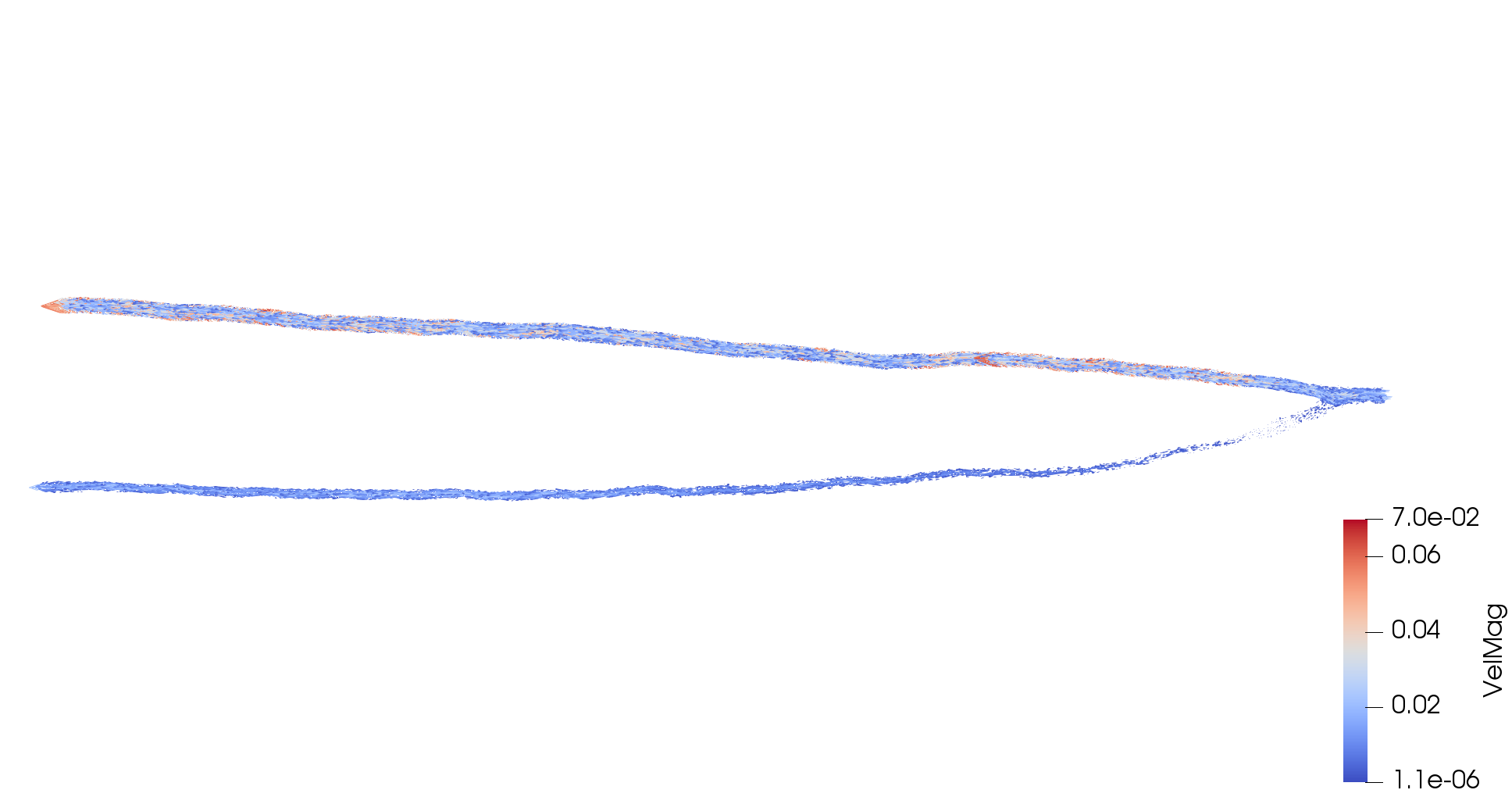} 
  \caption{Velocity - Elastic wall model}
  \label{Vel_F025}
 \end{subfigure}
  \begin{subfigure}{0.49\textwidth}
   \centering
   \includegraphics[width=0.98\textwidth]{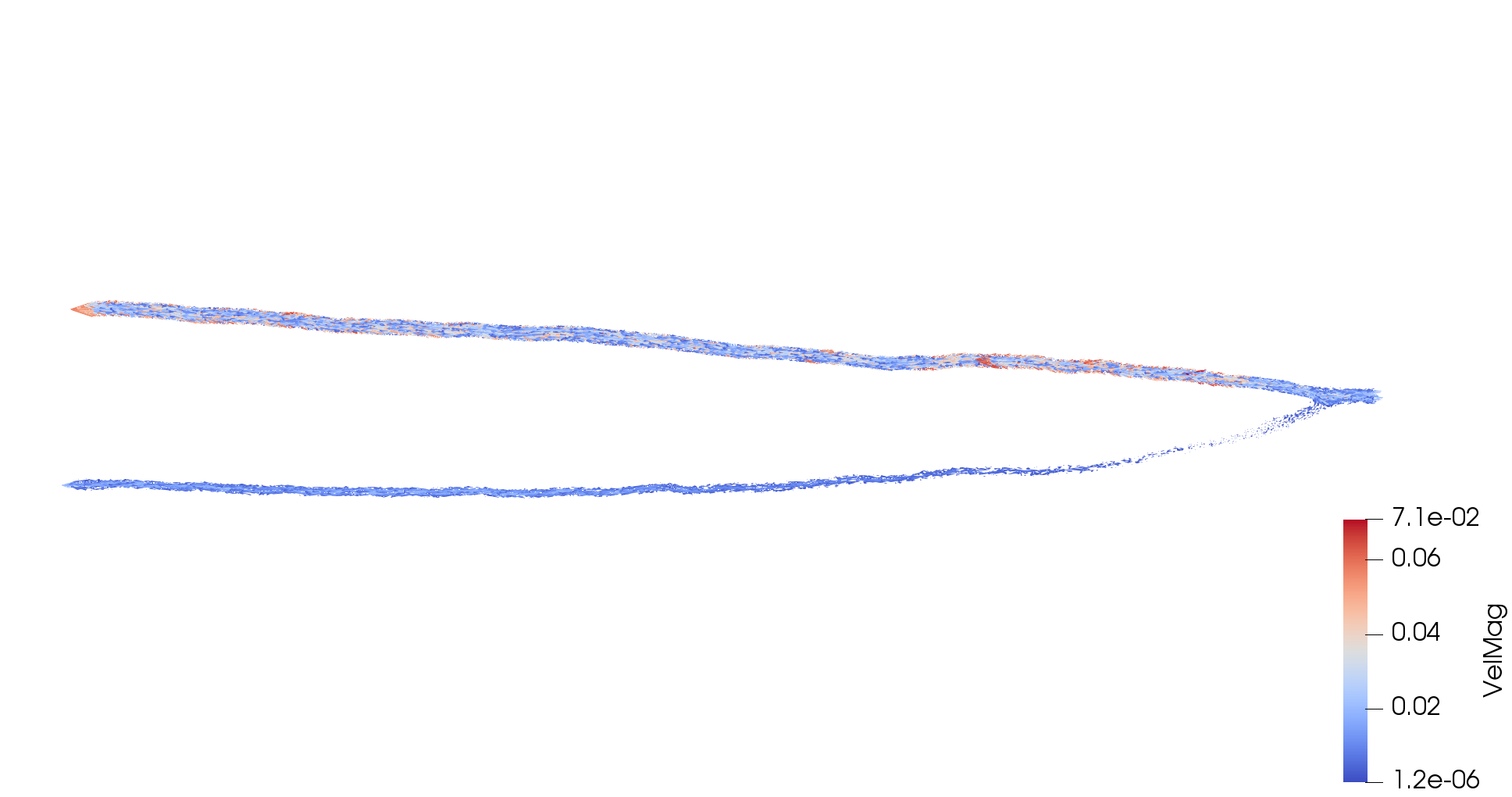} 
  \caption{Velocity - Rigid wall model}
  \label{Vel_noe}
 \end{subfigure}
 \begin{subfigure}{0.49\textwidth}
   \centering
   \includegraphics[width=0.98\textwidth]{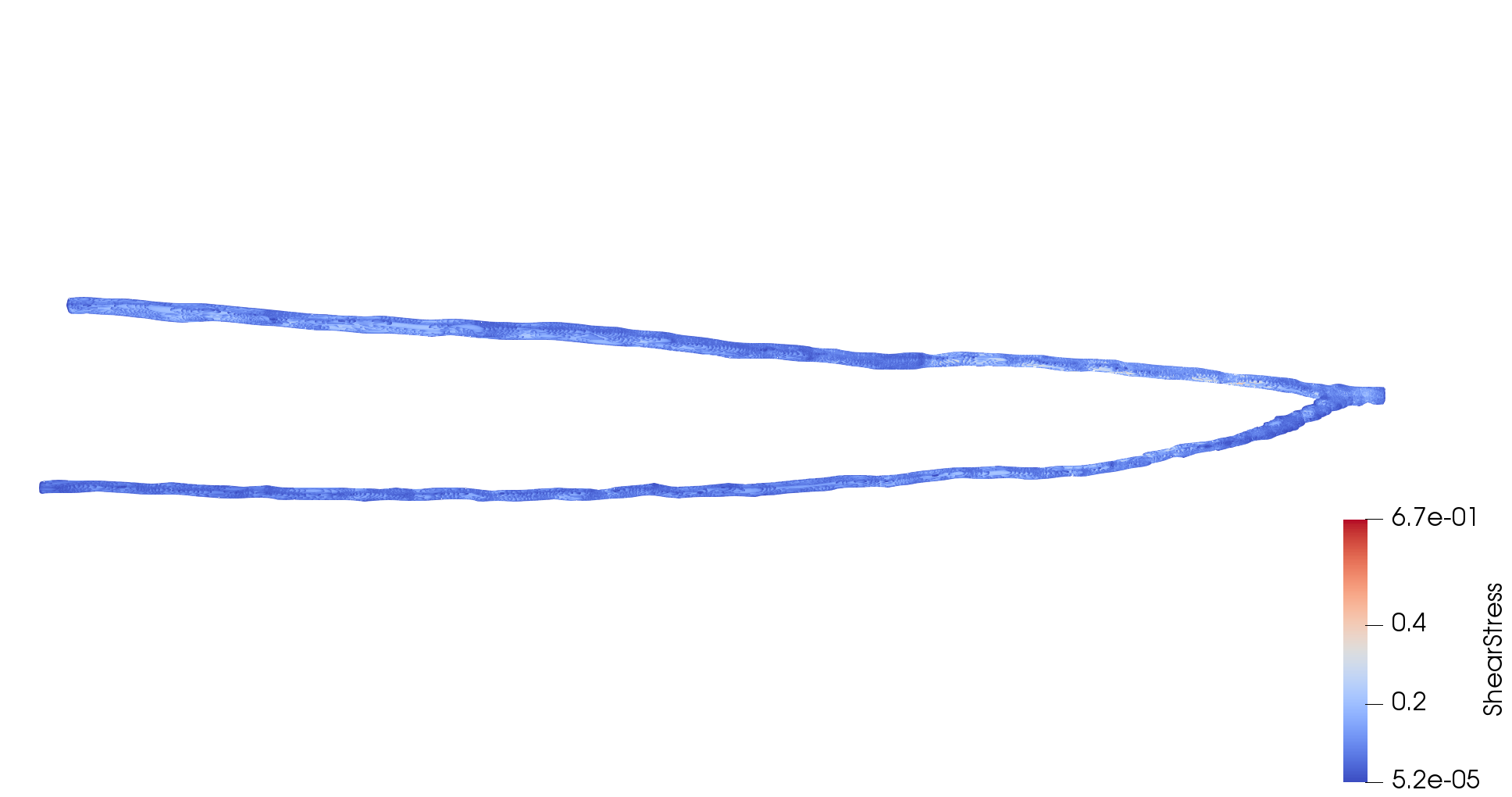} 
  \caption{Wall shear stress - Elastic wall model}
  \label{SS_F025}
 \end{subfigure}
  \begin{subfigure}{0.49\textwidth}
   \centering
   \includegraphics[width=0.98\textwidth]{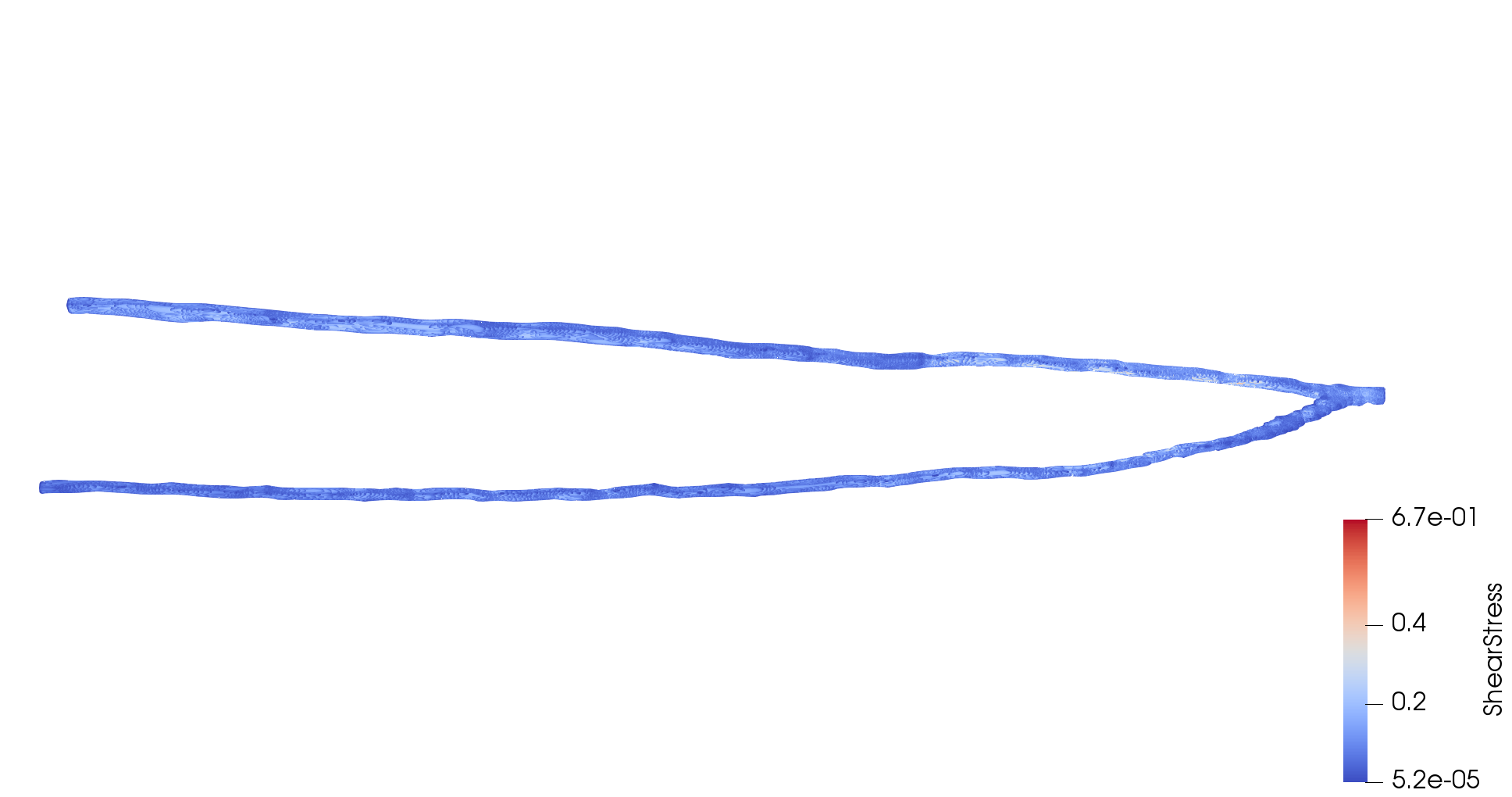} 
  \caption{Wall shear stress - Rigid wall model}
  \label{SS_noe}
 \end{subfigure}
  \caption{Velocity and wall shear stress fields for the arterial domain in their original dimensions}
  \label{ArteriesOriginal}
\end{figure}

\begin{figure}[!ht]
 \begin{subfigure}{0.49\textwidth}
   \centering
   \includegraphics[width=0.98\textwidth]{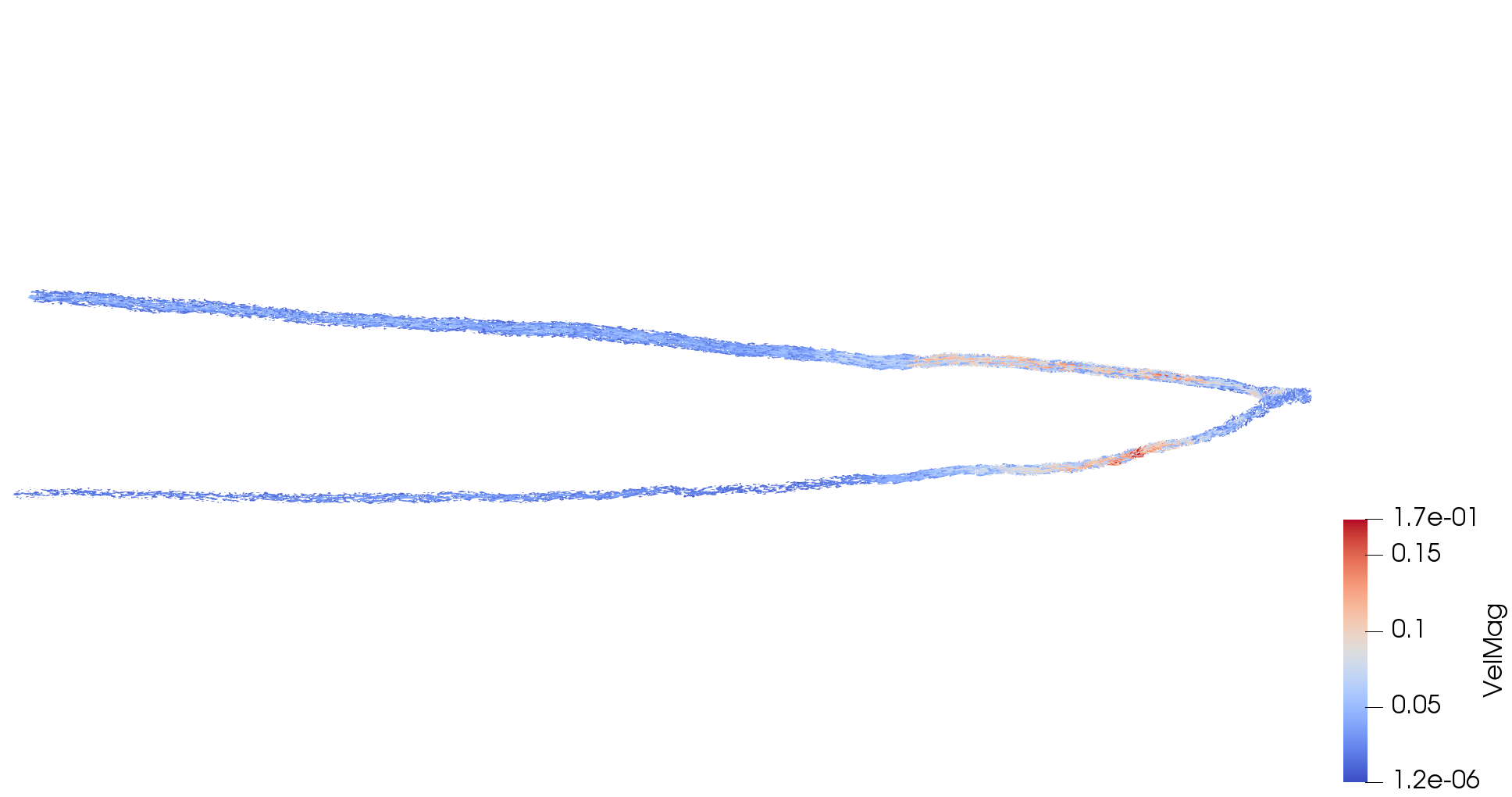} 
  \caption{Velocity - Elastic wall model}
  \label{DVel_F50}
 \end{subfigure}
  \begin{subfigure}{0.49\textwidth}
   \centering
   \includegraphics[width=0.98\textwidth]{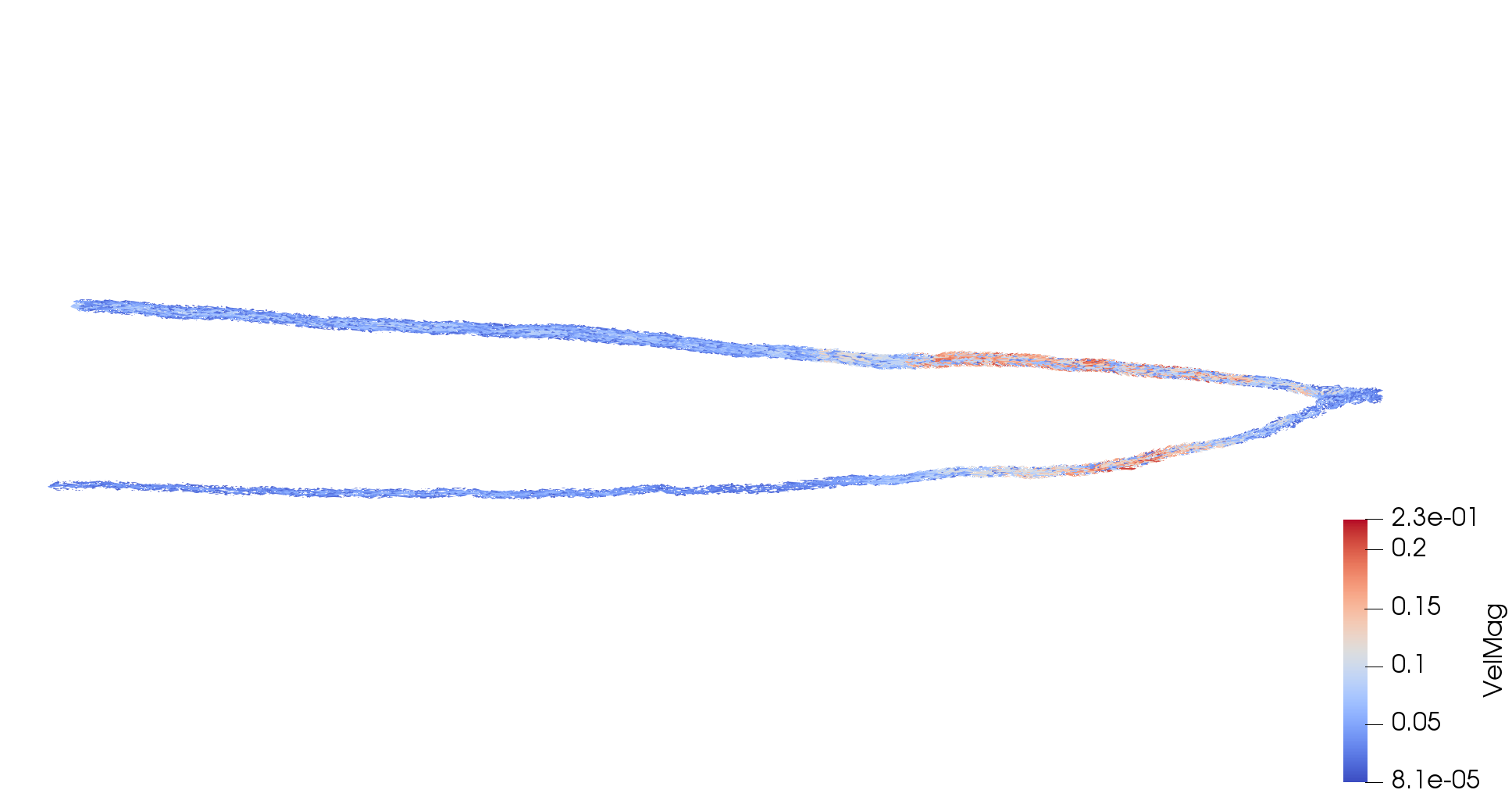} 
  \caption{Velocity - Rigid wall model}
  \label{DVel_noe}
 \end{subfigure}
 \begin{subfigure}{0.49\textwidth}
   \centering
   \includegraphics[width=0.98\textwidth]{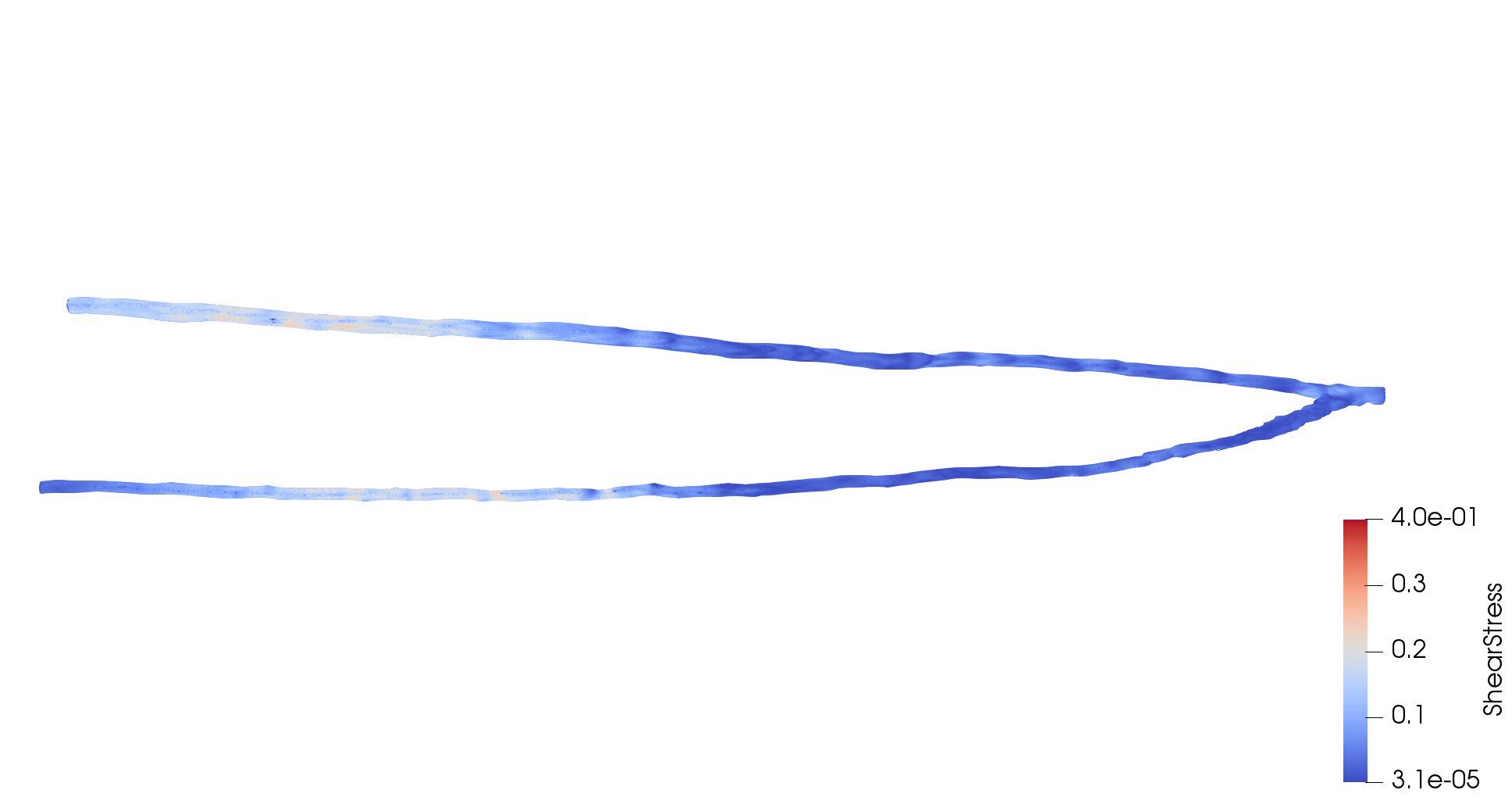} 
  \caption{Wall shear stress - Elastic wall model}
  \label{DSS_F50}
 \end{subfigure}
  \begin{subfigure}{0.49\textwidth}
   \centering
   \includegraphics[width=0.98\textwidth]{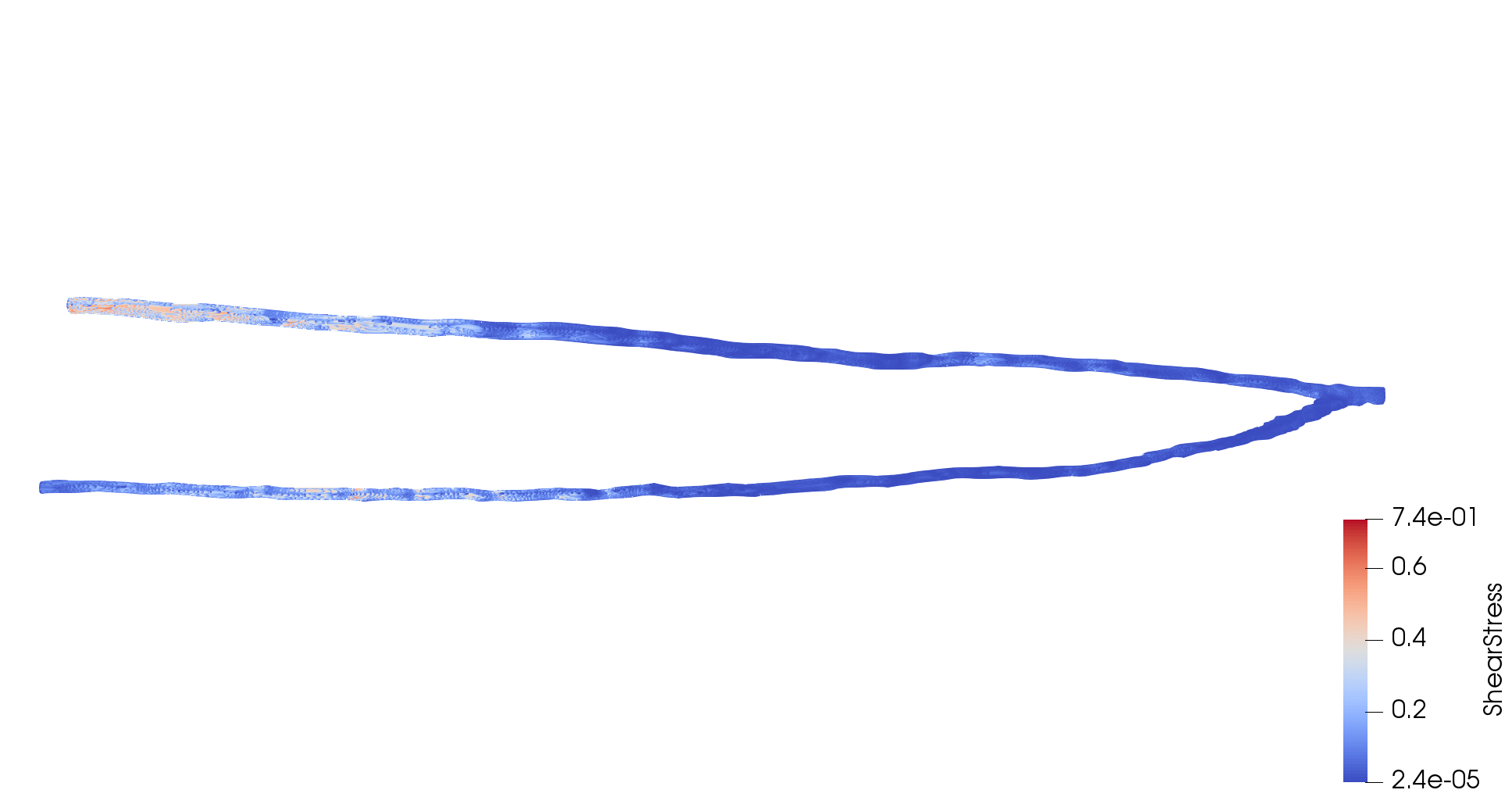} 
  \caption{Wall shear stress - Rigid wall model}
  \label{DSS_noe}
 \end{subfigure}
  \caption{Velocity and wall shear stress fields for the arterial domain in their dilated dimensions}
  \label{ArteriesDilated}
\end{figure}

\begin{figure}[!ht]
 \begin{subfigure}{0.49\textwidth}
   \centering
   \includegraphics[width=0.98\textwidth]{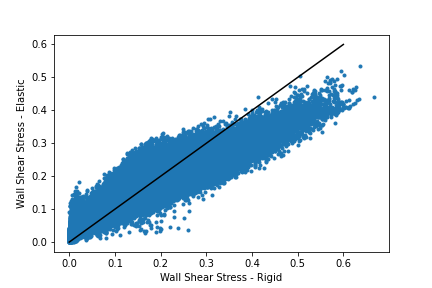} 
  \caption{Original dimensions}
  \label{ShearCompare}
 \end{subfigure}
  \begin{subfigure}{0.49\textwidth}
   \centering
   \includegraphics[width=0.98\textwidth]{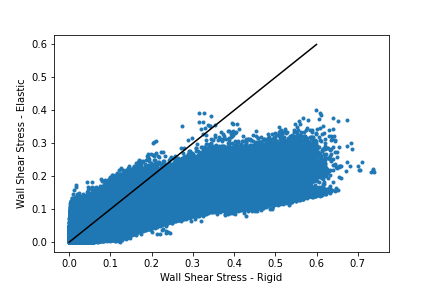} 
  \caption{Dilated dimensions}
  \label{DilateShearCompare}
 \end{subfigure}
  \caption{Pointwise comparison of instantaneous wall shear stress seen across the full domain of the arterial geometry in the rigid and elastic walled cases. The shape of the plot - generally recording higher wall shear stress in the rigid walled case - is consistent with that generated in other work using a fully coupled elastic wall model \cite{McGah2014}. The solid line represents the identity line.}
  \label{ShearComparison}
\end{figure}

In this section we demonstrate the behaviour of our model in patient specific vessels. The domain we are studying consists of the radial and ulnar arteries of the left forearm where we provide a velocity profile to the inlet plane (see Figure \ref{ArteriesLayout}). Fixed pressure conditions \cite{Nash2014} were applied to the outlets. Based on the geometry of the vessels we have chosen a value of $F$ of 0.025. To replicate larger vessels we have taken the same domain and adjusted the size of the lattice spacing to dilate the vessels by a factor of approximately four. Whilst not representative of a particular vessel it carries the characteristics of a patient-specific geometry and allows us to present results representative of a more flexible vascular domain. In this second case we have used a value of $F$ of 0.5. Figures \ref{ArteriesOriginal} and \ref{ArteriesDilated} illustrate the velocity and wall shear stress fields observed approximately 60\% through the simulation time. In both cases the lower shear stress observed in the elastic wall cases is consistent with observations made in other numerical studies of patient specific vessels \cite{McGah2014,Reymond2013}. This is made more explicitly clear when we compare the local, instantaneous wall shear stress between the rigid and elastic cases in Figure \ref{ShearComparison} where the shape of our `Limits of Agreement` plot of instantaneous wall shear stress is very similar to that presented in \cite{McGah2014} for time-averaged wall shear stress. This again demonstrates that our model is able to effectively capture behaviour expected from a fully coupled elastic wall model. \\

\section{Discussion}
In this paper we have presented a boundary condition that allows key features of elastic walled flow such as velocity profiles near walls and wall shear stress variations to be captured without the need to implement a complex computational coupling with a solid mechanics solver. This is achieved by applying a slip velocity at the wall of the domain that represents the flow at that physical location if an elastic wall was extended beyond it. This is calculated through a scaling parameter that can be estimated based on the physical properties of the simulated vessel and the expressions for Womersley flow in an elastic cylinder. Whilst our model does not, and is not intended to, perfectly capture the analytic flow profiles expected within an elastic walled cylinder, it is significantly more accurate than compared to results found using a rigid wall assumption. This was achieved with no loss of computational performance compared to a commonly used boundary condition for rigid walls in the lattice Boltzmann method. This indicates that the use of our model would represent an effective return on the investment of implementing it within a simulation. Although we have discussed $F$ as a global parameter is this paper, there is no fundamental reason why it could not be tuned locally within a geometry of widely varying vessel diameters such as a whole human vascular tree.  \\

When compared to other sources of uncertainty related to measuring and validating blood flow in personalised geometries  - particularly those associated with clinical measurement techniques such as ultrasound or MRI, the error indicated by our simple model is of a similar order of magnitude. For example, \cite{Merkx2013,Merkx2013a} record a significant difference between the diameters recorded using an MRI technique and ultrasound and state that the MRI technique may overestimate vessel diameters by 38\% potentially in part due to the sensitivity of the technique to patient position. Furthermore, \cite{Merkx2013a} notes that for the radial artery may vary diameter 4-7\% in day-to-day function. Image analysis techniques used to assess medical images can generate similar levels of variation and uncertainty \cite{Kaufhold2018}. The uncertainty in velocity measurements from MRI is further discussed in \cite{Bruschewski2016}. Indeed, \cite{Brindise2019} summarises the challenge in validating against MRI derived data as ``A major challenge for any multi-modality study that uses \textit{in vivo} measurements is that no `ground-truth’ flow field can be established''. Keeping these factors in mind, the errors presented from our model compared to those generated with a rigid wall approximation seem acceptable. In spite of the comments from \cite{Brindise2019}, further validation of our model could be achieved through comparison to flow fields in a personalised vessel with known geometries and properties. \\ 

Further development of this model would be best focussed on how its implementation could be improved to effectively study domains with a greater spread in vascular diameters and resolution whilst retaining the locality of the implementation. As noted above this could be achieved with a local specification of the $F$ parameter. How a global (or regional) value of $F$ could be better tuned to different flow scenarios would also be of interest to the study of large-scale vascular structures. \\

With a view towards the development of a virtual human, the use of our model would permit efficient deployment of high resolution, 3D blood flow simulation with the effect of elastic walls included. Not having to support an explicit coupling for the solid mechanics of the vessel walls will reduce the communication burden of the simulation and allow resources to be deployed to other components of a virtual human model.

\section*{Funding}
We acknowledge funding support from European Commission CompBioMed Centre of Excellence (Grant No. 675451 and 823712). Support from the UK Engineering and Physical Sciences Research Council under the project `UK Consortium on Mesoscale Engineering Sciences (UKCOMES)' (Grant No. EP/R029598/1) is gratefully acknowledged. We acknowledge funding support from MRC for a Medical Bioinformatics grant (MR/L016311/1), and special funding from the UCL Provost. \\

The authors gratefully acknowledge the Gauss Centre for Supercomputing e.V. (\url{www.gauss-centre.eu}) for funding this project by providing computing time on the GCS Supercomputer SuperMUC-NG at Leibniz Supercomputing Centre (\url{www.lrz.de}). 

\bibliographystyle{unsrtnat_JMmod}
\bibliography{2020_ElasticWallsPaper}

\newpage

\section*{Appendix}

\setcounter{table}{0}
\renewcommand{\thetable}{A\arabic{table}}

\begin{table}[!htbp]
\caption{Simulation Parameters}
\begin{center}
\begin{tabular}{|c|p{2cm}|p{2cm}|p{2cm}|p{2cm}|p{2cm}|}
\hline 
\textbf{Test} & Cylinder ($R$=50$\Delta x$) & Cylinder ($R$=100$\Delta x$) & Cylinder ($R$=200$\Delta x$) & Arteries - Original & Arteries - Dilated \\ 
\hline  
\textbf{$\Delta x$ [m]} & 6.0e-5 & 3.0e-5 & 1.5e-5 & 5.0e-5 &  2.1e-04 \\  
\hline 
\textbf{$\Delta t$ [s]} & 1.10e-05 & 5.48e-06 & 2.74e-06 & 5.0e-6 &  1.0e-4 \\  
\hline 
\textbf{$\tau$} & 0.537 & 0.573 & 0.646 & 0.527 & 0.527 \\  
\hline 
\textbf{Steps} & 1000000 & 2000000 & 4000000 & 660000 & 33000  \\  
\hline 
\textbf{Lattice sites} & 5,195,466 & 41,596,265 & 332,841,364 & 6,128,855 &  6,128,855  \\  
\hline
\textbf{Cores} & 3072 & 6000 & 12000 & 2400 & 2400 \\  
\hline 
\end{tabular} 
\end{center}
\label{tab:SimValues}
\end{table}

All simulations were run on the SuperMUC-NG supercomputer (\url{https://doku.lrz.de/display/PUBLIC/SuperMUC-NG}) situated at the Leibniz Supercomputing Centre, Germany. This machine uses Intel Skylake processore (Xeon Platinum 8174) with 48 CPU cores per node. Simulations were run using the full complement of cores on each node. Nodes are connected with an OmniPath interconnect configured in an island layout. HemeLB was compiled using the default 2019 versions of Intel C++ compilers and MPI. \\

\noindent The version of HemeLB used for this study can be obtained from \url{https://github.com/UCL-CCS/HemePure}. 

\end{document}